%% file: shapiro_main2e_final.tex
\documentclass[12pt]{book}
\usepackage{myplenum}
\usepackage{graphicx}
\begin{document}
\input shapiro_body_final.tex

\end{document}

%% file: shapiro_body_final.tex
\chapter{LONG-DISTANCE HIGH-FIDELITY TELEPORTATION USING SINGLET
STATES}

\author{Jeffrey H. Shapiro}

\affiliation{Department of Electrical Engineering and Computer Science\\
Research Laboratory of Electronics\\
Laboratory for Information and Decision Systems\\
Massachusetts Institute of Technology\\
Cambridge, MA 02139}

\abstract{A quantum communication system is proposed that uses
polarization-entangled photons and trapped-atom quantum memories.  This
system is capable of
long-distance, high-fidelity teleportation, and long-duration quantum storage.}

\section{INTRODUCTION}

The preeminent obstacle to the development of quantum information
technology is the difficulty
of transmitting quantum information over noisy and lossy quantum
communication channels,
recovering and refreshing the quantum information that is received, and
then storing it in a
reliable quantum memory.  This paper proposes a singlet-based approach to
quantum communication that
uses a novel ultrabright narrowband source of polarization-entangled photon
pairs,\refnote{\cite{ShapiroWong}} and a trapped-atom quantum
memory\refnote{\cite{LloydShahriarHemmer}} whose loading can be
nondestructively verified and whose
structure permits all four Bell-state measurements to be performed.  The
system is designed to operate
with standard telecommunication fiber as its transmission medium.  It can
achieve a
loss-limited throughput as high as 200 entangled-pairs/sec with a 97.5\%
fidelity over a 50\,km path
when there is 10\,dB of fixed loss in the overall system and 0.2\,dB/km
propagation loss in the fiber.
This long-distance high-fidelity quantum transmission is accomplished
without the use of entanglement
swapping, i.e., no quantum repeaters, and without the use of entanglement
purification or quantum error
correction.

\section{TELEPORTATION USING SINGLET STATES}
The notion that singlet states could be used to achieve teleportation is
due to Bennett {\it et
al}.\refnote{\cite{Bennett}}  The transmitter and receiver stations share
the entangled qubits of a
singlet state, $|\psi\rangle_{\mbox{\scriptsize TR}} =
\left(|0\rangle_{\mbox{\scriptsize T}}|1\rangle_{\mbox{\scriptsize R}} -
|1\rangle_{\mbox{\scriptsize T}}|0\rangle_{\mbox{\scriptsize
R}}\right)/\sqrt{2}.$
The transmitter accepts an input-mode qubit,
$|\Psi\rangle_{\mbox{\scriptsize in}} =
\alpha|0\rangle_{\mbox{\scriptsize in}} + \beta|1\rangle_{\mbox{\scriptsize
in}}$, leaving the
input-mode, transmitter, and receiver in the joint state
$|\Psi\rangle_{\mbox{\scriptsize
in}}|\psi\rangle_{\mbox{\scriptsize TR}}$.
Making the Bell-state measurements,  $\{\left(|1\rangle_{\mbox{\scriptsize
in}}|0\rangle_{\mbox{\scriptsize T}} \pm  |0\rangle_{\mbox{\scriptsize
in}}|1\rangle_{\mbox{\scriptsize
T}}\right)/2,$ $\left(|1\rangle_{\mbox{\scriptsize
in}}|1\rangle_{\mbox{\scriptsize T}} \pm
|0\rangle_{\mbox{\scriptsize in}}|0\rangle_{\mbox{\scriptsize
T}}\right)/2\}$, on the joint
input-mode/transmitter system then yields the two bits of classical
information that
the receiver needs to reconstruct the input state, i.e., to complete the
teleportation process.  An
initial experimental demonstration of teleportation using singlet states
was performed by the Innsbruck
group.\refnote{\cite{Bouwmeester1},\cite{Bouwmeester2}}  There were several
significant limitations to
this initial demonstration, however, which preclude its forming the basis
for useful long-distance
quantum-state communication.  First, because only one of the Bell states
was measured, the
demonstration was conditional:  teleportation only occurred when the
input-mode/transmitter state
projected onto that Bell state.  Second, the demonstration was a table-top
experiment: there was no
provision for long-distance transmission.  Finally, the demonstration did
not include a
quantum memory:  the teleported state could not be stored for application to
quantum cryptography or quantum computation.  In the next section we
outline our
proposal for a singlet-based quantum communication system that remedies all
of these
limitations.

\section{LONG-DISTANCE TELEPORTATION WITH QUANTUM MEMORY}
Consider the quantum communication system shown in Fig.~1.
\begin{figure}[htb]
\vskip .2cm
\begin{center}
\includegraphics[width=.5\hsize]{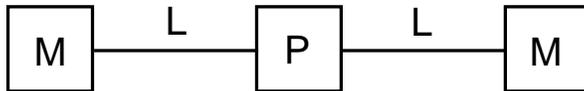}
\end{center}
\vskip -.25cm
\caption{Schematic of long-distance quantum communication system:  $P$ =
ultrabright
narrowband source of polarization-entangled photon pairs; $L$ = $L$\,km of
standard
telecommunication fiber; $M$ = trapped-atom quantum memory.}
\end{figure}
An ultrabright narrowband source of polarization-entangled photon
pairs\refnote{\cite{ShapiroWong}}
launches  the entangled qubits from a singlet state into two $L$-km-long
standard telecommunication fibers.  The photons emerging from the fibers
are then loaded into
trapped-atom quantum memories.\refnote{\cite{LloydShahriarHemmer}}  These
memories store the
photon-polarization qubits in long-lived hyperfine
levels.  Because it is compatible with fiber-optic transmission, this
configuration is capable of
long-distance teleportation.  Because of the long decoherence
times that can be realized with trapped atoms, this configuration supports
long-duration quantum
storage.   We devote the rest of this section to summarizing the basic
features of our proposal.

Each $M$ block in Fig.~1 is a quantum memory in which a single ultra-cold
$^{87}$Rb atom ($\sim6$\,MHz
linewidth) is confined by a CO$_2$-laser trap in an ultra-high vacuum
chamber with cryogenic walls
within a high-finesse ($\sim15$\,MHz linewidth) single-ended optical
cavity.  This memory can absorb
a 795\,nm photon, in an arbitrary polarization state, transferring the
qubit from the photon to
the degenerate $B$ levels of Fig.~2a and thence to long-lived storage
levels, by coherently driving the
$B$-to-$D$ transitions.  (We are using abstract symbols here for the hyperfine
levels of rubidium, see Ref.~\cite{LloydShahriarHemmer} for the actual
atomic levels involved as
well as a complete description of the memory and its operation.)  With a
liquid helium cryostat, so that
the background pressure is less than 10$^{-14}$\,Torr, the expected
lifetime of the trapped rubidium
atom will be more than an hour.  Moreover, the decoherence time can be
expected to be about the same as
this lifetime for the levels we have chosen to use for storage.  By using
optically off-resonant Raman
(OOR) transitions,  the Bell states of two atoms in a single vacuum-chamber
trap can be converted to
superposition states of one of the atoms. All four Bell measurements can
then be made, sequentially,
by detecting the presence (or absence) of fluorescence as an appropriate
sequence of OOR laser
pulses is applied to the latter atom.  The Bell-measurement results (two
bits of classical information) in one memory can be sent to a distant
memory, where
(at most) two additional OOR pulses are needed to complete the Bennett {\it
et al}.\ state
transformation.  The qubit stored in a trapped rubidium atom can be
converted back into a photon---with the same polarization information as
the one whose qubit
was stored---by reversing the Raman excitation process that occurs during
memory loading.
\begin{figure}[htb]
\begin{center}
\includegraphics[width=.9\hsize]{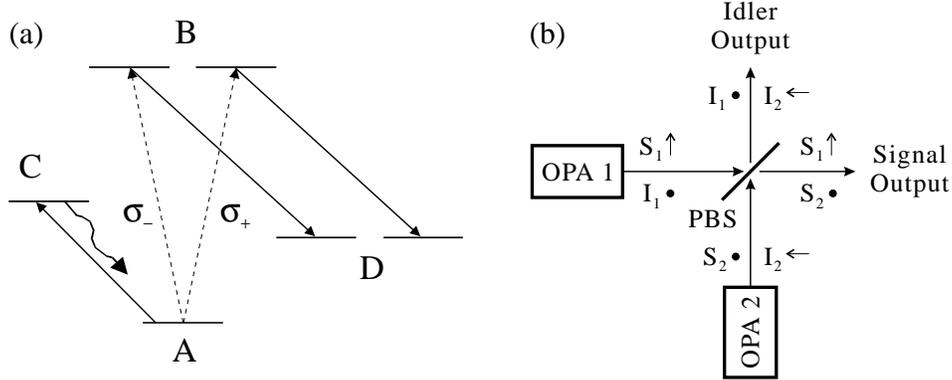}
\end{center}
\vskip -.25cm
\caption{Essential components of singlet-state quantum communication system
from Fig.~1.
Left (a), simplified atomic-level schematic of the trapped rubidium atom
quantum memory:  $A$-to-$B$
transition occurs when one photon from an entangled pair is absorbed;
$B$-to-$D$ transition is
coherently driven to enable storage in the long-lived $D$ levels;
$A$-to-$C$ cycling transition
permits nondestructive determination of when a photon has been absorbed.  Right
(b), ultrabright narrowband source of polarization-entangled photon pairs:
each optical
parametric amplifier (OPA1 and OPA2) is type-II phase matched; for each
optical beam the
propagation direction is $\hat{z}$, and $\hat{x}$ and $\hat{y}$
polarizations are denoted by
arrows and bullets, respectively; PBS, polarizing beam splitter.}
\end{figure}

The
$P$-block in Fig.~1 is an ultrabright narrowband source of
polarization-entangled photon
pairs, capable of producing $\sim10^{6}$ pairs/sec in $\sim30$\,MHz
bandwidth by appropriately
combining the signal and idler output beams from two doubly-resonant
type-II phase-matched
optical parametric amplifiers (OPAs), as sketched in
Fig.~2b.\refnote{\cite{ShapiroWong}}
The Innsbruck teleportation experiment used parametric downconversion as its
source of polarization-entangled photon pairs.  This process is
intrinsically very broadband
($\sim10^{13}\,$Hz bandwidth), whereas a trapped-atom quantum memory only
absorbs photons within a very
narrow ($\sim10^{7}\,$Hz) bandwidth.  As a result, the brightest
downconverter source
reported to date\refnote{\cite{Kwiat}} might only produce
$\sim15$\,pairs/sec in this narrow optical
bandwidth.  This is why an ultrabright narrowband source is so essential.
Our source has the following
properties.  The fluorescence spectrum of the signal and idler beams is
controlled by the
doubly-resonant OPA cavities.  These can be advantageously and easily
tailored to produce the desired
(factor-of-two broader than the memory-cavity's) bandwidth.  By using
periodically-poled potassium
titanyl phosphate (PPKTP), a quasi-phase-matched type-II nonlinear
material, we can produce
$\sim10^{6}$\,pairs/sec at the 795\,nm wavelength of the rubidium memory
for direct memory-loading
(i.e., local-storage) applications.  For long-distance transmission to
remotely-located memories, we use
a different PPKTP crystal and pump wavelength to generate
$\sim10^{6}$\,pairs/sec in the
1.55-$\mu$m-wavelength low-loss fiber transmission window. After fiber
propagation we then shift
the entanglement to the 795\,nm wavelength needed for the rubidium-atom
memory via quantum-state
frequency translation, a procedure previously proposed and demonstrated by
Kumar\refnote{\cite{Kumar},\cite{HuangKumar}} and shown schematically in
Fig.~3.
\begin{figure}[htb]
\vskip 0.35cm
\begin{center}
\includegraphics[width=.75\hsize]{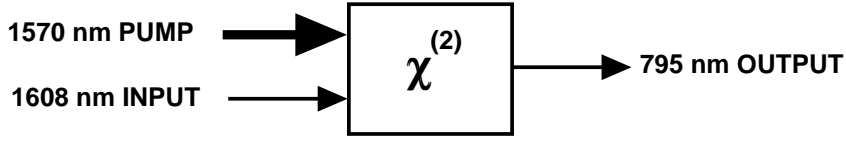}
\end{center}
\vskip -0.5cm
\caption{Schematic of quantum frequency conversion:  a strong pump beam at
1570\,nm converts a qubit
photon received at 1608\,nm (in the low-loss fiber transmission window) to
a qubit photon at the
795\,nm wavelength of the $^{87}$Rb quantum memory.}
\end{figure}

Transmission in the
1.55-$\mu$m-wavelength window is not enough to make singlet-based quantum
communication fully compatible
with standard telecommunication fiber.  It is also crucial to ensure that
polarization is not degraded by the propagation process.  Our scheme for
polarization
maintenance, shown schematically in Fig.~4, relies on time-division
multiplexing. Time slices from the
signal beams from our two OPAs are sent down one fiber in the same linear
polarization but in
nonoverlapping time slots, accompanied by a strong out-of-band laser pulse.
By tracking and restoring
the linear polarization of the strong pulse, we can restore the linear
polarization of the signal-beam
time slices at the far end of the fiber.  After this linear-polarization
restoration,
we then reassemble a time-epoch of the full vector signal beam by delaying
the first time
slot and combining it on a polarizing beam splitter (PBS) with the second
time slot
after the latter has had its linear polarization rotated  by 90$^\circ$.  A
similar
procedure is performed to reassemble idler time-slices after they have
propagated down the
other fiber in Fig.~1.  In effect, this replaces the source-located passive
PBS in Fig.~2b with a
time-gated memory-located polarization combiner at the far end
of each fiber.  This approach, which is inspired by the Bergman {\it et
al}.\ two-pulse
fiber-squeezing experiment,\refnote{\cite{Bergman}} common-modes out the
vast majority of the phase
fluctuations and the polarization birefringence incurred in the fiber,
permitting standard
telecommunication fiber to be used in lieu of the lossier and much more
expensive
polarization-maintaining fiber.
\begin{figure}[htb]
\begin{center}
\includegraphics[width=.8\hsize]{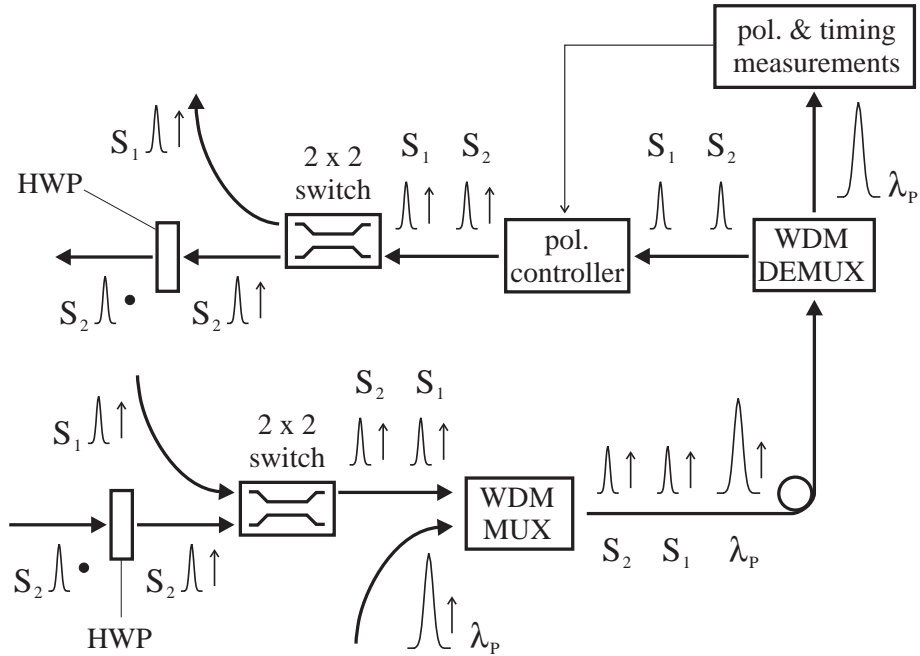}
\end{center}
\vskip -.25cm
\caption{Transmission of time-division multiplexed
signal beams from OPA1 and OPA2 through an optical fiber.  With the use of
a half-wave plate
(HWP) the signal pulses and the pilot pulse are linearly polarized the same
way.
The pilot pulse---at a wavelength $\lambda_p$ which is different from the
signal or idler
wavelengths---is injected into and extracted from the fiber using a
wavelength-division
multiplexer (WDM MUX) and a wavelength-division demultiplexer (WDM DEMUX),
respectively.  Not shown
is the final polarizing beam splitter for combining the two signal-beam
outputs, cf.\ Fig.~2b.  A
similar overall arrangement is used to transmit the idler beams from OPA1
and OPA2.}
\end{figure}

\section{LOSS-LIMITED PERFORMANCE ANALYSIS}
Quantum communication is carried out in the Fig.~1 configuration via the
following protocol.
The entire system is clocked.  Time slots of signal and idler (say 400\,ns
long) are
transmitted down optical fibers to the quantum memories.  These slots are
gated into the memory
cavities---with their respective atoms either physically displaced or
optically detuned so that
no
$A$-to-$B$ (i.e., no 795\,nm) absorptions occur.  After a short loading
interval (a few
cold-cavity lifetimes, say 400\,ns), each atom is moved (or tuned) into the
absorbing
position and $B$-to-$D$ coherent pumping is initiated.  After about
100\,ns, coherent pumping
ceases and the $A$-to-$C$ cycling transition (shown in Fig.~2a) is
repeatedly driven (say 30
times, taking nearly 1\,$\mu$s).  (To avoid
spontaneous decay from $B$, the two-step storage sequence can be compressed
into a single off-resonant Raman
transition.\refnote{\cite{Cirac},\cite{Dur}})  By monitoring a cavity
for the fluorescence from this cycling transition, we can reliably detect
whether or not a 795\,nm
photon has been absorbed by the atom in that cavity.  If neither atom or if
only one atom has absorbed
such a photon, then we cycle both atoms back to their $A$ states and start
anew.  If no
cycling-transition fluorescence is detected in either cavity, then, because
we have employed enough
cycles to ensure very high probability of detecting that the atom is in its
$A$ state, it must be that
both atoms have absorbed 795\,nm photons and stored the respective qubit
information
in their long-lived degenerate
$D$ levels.  These levels are not resonant with the laser driving the cycling
transition, and so the loading of our quantum memory is nondestructively
verified in this
manner.

We expect that the preceding memory-loading protocol can be run at rates as
high as $R =
500$\,kHz, i.e., we can get an independent try at loading an entangled
photon pair into
the two memory elements of Fig.~1 every 2\,$\mu$s.  With a high probability,
$P_{\mbox{\scriptsize erasure}}$, any particular memory-loading trial will
result in an
erasure, i.e., propagation loss and other inefficiencies combine to
preclude both atoms
from absorbing photons in the same time epoch.  With a small probability,
$P_{\mbox{\scriptsize success}}$, the two atoms will absorb the photons
from a single
polarization-entangled pair, viz., we have a memory-loading success.  With
a much
smaller probability, $P_{\mbox{\scriptsize error}}$, both atoms will have
absorbed photons but these photons will not have come from a single
polarization-entangled pair; this is the error event.  These three events
constitute
a complete taxonomy of loading-trial possibilities, i.e., their respective
probabilities sum
to unity.

In terms of $\{R, P_{\mbox{\scriptsize erasure}}, P_{\mbox{\scriptsize
success}},
P_{\mbox{\scriptsize error}}\}$ it is easy to identify the key
figures-of-merit for the
Fig.~1 configuration.  First, there is its robustness to propagation losses
and other
inefficiencies.  These effects merely
increase $P_{\mbox{\scriptsize erasure}}$ and hence reduce the throughput,
i.e., the number of
successful entanglement-loadings/sec, $N_{\mbox{\scriptsize success}} \equiv
RP_{\mbox{\scriptsize success}}$, that could be achieved if the quantum
memories each contained a
lattice of trapped atoms for sequential loading of many pairs.  It is the
loading errors, which
occur with probability
$P_{\mbox{\scriptsize error}}$, that provide the ultimate limit on the
entanglement
fidelity of the Fig.~1 configuration.  This loss-limited fidelity is given by
$F_{\mbox{\scriptsize max}} = 1-P_{\mbox{\scriptsize error}}/
2(P_{\mbox{\scriptsize success}} + P_{\mbox{\scriptsize error}})$,
where we have assumed that the error event loads independent,
randomly-polarized
photons into each memory.

\subsection{OPA Statistics}
To quantify the loss-limited throughput and entanglement fidelity of the
Fig.~1 system we must first
quantify the behavior of the dual-OPA source.  We begin by showing that
this source---under ideal
lossless conditions---does indeed produce the desired singlet state.
Assume matched signal and idler
cavities, each with linewidth
$\Gamma$, zero detuning, and no excess loss. Also assume anti-phased
pumping at a fraction, $G^2$, of
oscillation threshold, with no pump depletion or excess noise.  From
Ref.~\cite{ShapiroWong} we then
have that the  output beams from OPAs 1 and 2 are in an entangled,
zero-mean Gaussian pure state, which
is completely characterized by the following normally-ordered and
phase-sensitive correlation
functions:
\begin{eqnarray}
\langle \hat{A}_{k_j}^{\dagger}(t+\tau)\hat{A}_{k_j}(t)\rangle &=&
\frac{\displaystyle G\Gamma}{\displaystyle 2}\!\left[
\frac{\displaystyle \exp[-(1-G)\Gamma|\tau|]}{\displaystyle 1-G} -
\frac{\displaystyle \exp[-(1+G)\Gamma|\tau|]}{\displaystyle 1+G} \right],
\label{normalorder}\\
\langle \hat{A}_{S_j}(t+\tau)\hat{A}_{I_j}(t)\rangle &=&
\frac{\displaystyle (-1)^{j-1}G\Gamma}{\displaystyle 2}\!\left[
\frac{\displaystyle \exp[-(1-G)\Gamma|\tau|]}{\displaystyle 1-G} +
\frac{\displaystyle \exp[-(1+G)\Gamma|\tau|]}{\displaystyle 1+G} \right],
\label{phasesens}
\end{eqnarray}
where $\{\,\hat{A}_{k_j}(t)e^{-i\omega_kt}: k = S\mbox{ (signal)}, I\mbox{
(idler)}, j = 1,2\,\}$
are positive-frequency, photon-units OPA-output field operators.

After combining the outputs of OPAs 1 and 2 into vector fields
$\hat{\vec{A}}_{S}(t)$ and
$\hat{\vec{A}}_{I}(t)$, we can show that the Fourier component of the
vector-signal field at
frequency $\omega_S + \Delta\omega$ and the vector-idler Fourier component
at frequency $\omega_I
-\Delta\omega$ are in the entangled Bose-Einstein state,
\begin{equation}
|\psi\rangle_{\rm SI} = \sum_{n = 0}^{\infty}
\sqrt{\frac{\displaystyle \bar{N}^n}{\displaystyle (\bar{N}+1)^{n+1}}}
|n\rangle_{S_x}|n\rangle_{I_y}
\sum_{n = 0}^{\infty}
(-1)^{n}\sqrt{\frac{\displaystyle \bar{N}^n}{\displaystyle (\bar{N}+1)^{n+1}}}
|n\rangle_{S_y}|n\rangle_{I_x}
\end{equation}
in number-ket representation, where $\bar{N} =
4G^2/[(1-G^2-\Delta\omega^2/\Gamma^2)^2 +
4\Delta\omega^2/\Gamma^2]$ is the average photon number per mode at
detuning $\Delta\omega$.  For
$\bar{N}\ll 1$, this joint state reduces to,
\begin{equation}
|\psi\rangle_{\rm SI} \approx
\frac{1}{\displaystyle \bar{N} + 1}|0\rangle_{S_x}|0\rangle_{I_y}
|0\rangle_{S_y}|0\rangle_{I_x}\nonumber +
\sqrt{\frac{\displaystyle \bar{N}}{\displaystyle (\bar{N} + 1)^3}}
(|1\rangle_{S_x}|1\rangle_{I_y}|0\rangle_{S_y}|0\rangle_{I_x}
-|0\rangle_{S_x}|0\rangle_{I_y}|1\rangle_{S_y}|1\rangle_{I_x}),
\end{equation}
i.e., it is predominantly vacuum, augmented by a small amount of the
singlet state.

The presence of excess loss within the OPA cavities, and/or propagation
loss along the fiber can be
incorporated into this OPA analysis in a straightforward
manner.\refnote{\cite{WongLeongShapiro}}
Assuming symmetric operation, in which the signal and idler encounter
identical intracavity and
fiber losses, then the correlation-function formulas,
Eqs.~\ref{normalorder} and
\ref{phasesens}, are merely multiplied by $\eta_L\gamma/\Gamma$, where
$\eta_L<1$ is the transmission
through the fiber ($\eta_L = 10^{0.02L}$ for $L$\,km of 0.2\,dB/km-loss
fiber), and $\gamma<\Gamma$ is
the output-coupling rate of the OPA cavity.

\subsection{Cavity-Loading Statistics}
To analyze our cold-cavity loading protocol, we relate
the annihilation operators of the internal cavity modes---over the
$T_c$-sec-long loading interval---to
the incoming signal and idler field operators as follows:
\begin{equation}
\hat{\vec{a}}_{k}(T_c) = \hat{\vec{a}}_{k}(0)e^{-\Gamma_cT_c} +
\int_{0}^{T_c}\!dt\,e^{-\Gamma_c(T_c-t)}\left[\sqrt{2\gamma_c}\hat{\vec{A}}_{k}(
t)
+\sqrt{2(\Gamma_c-\gamma_c)}\hat{\vec{A}}_{k_v}(t)\right],
\end{equation}
for $k=S,I$,
where $\gamma_c<\Gamma_c$ is the input-coupling rate and
$\Gamma_c$ is the linewidth of the (assumed to be identical for signal and
idler) memory cavities.  The
initial intracavity operators and the loss-operators,
$\{\hat{\vec{a}}_{k}(0),\hat{\vec{A}}_{k_v}(t)\}$, are in vacuum states.

It is now easy to show that the joint density operator (state) for
$\{\hat{\vec{a}}_{S}(T_c),\hat{\vec{a}}_{I}(T_c)\}$, takes the factored form,
$\hat{\rho}_{\vec{S}\vec{I}} =
\hat{\rho}_{S_xI_y}
\hat{\rho}_{S_yI_x}$, where the two-mode density operators on the
right-hand side are Gaussian mixed
states given by the anti-normally ordered characteristic functions,
\begin{eqnarray}
\mbox{tr}\!\left[\hat{\rho}_{S_xI_y} e^{-\zeta_{S}^*\hat{a}_{S_x} -
\zeta_{I}^*\hat{a}_{I_y}}
e^{\zeta_{S}\hat{a}_{S_x}^\dagger +
\zeta_{I}\hat{a}_{I_y}^\dagger}\right] &=&
\mbox{tr}\!\left[\hat{\rho}_{S_yI_x} e^{-\zeta_{S}^*\hat{a}_{S_y} +
\zeta_{I}^*\hat{a}_{I_x}}
e^{\zeta_{S}\hat{a}_{S_y}^\dagger -
\zeta_{I}\hat{a}_{I_x}^\dagger}\right]\\
&=& \exp\!\left[-(1+\bar{n})(|\zeta_{S}|^2 + |\zeta_{I}|^2)
+2\tilde{n}\mbox{Re}(\zeta_{S}\zeta_{I})\right],
\end{eqnarray}
where $\bar{n} \equiv I_{-} - I_{+}$ and $\tilde{n} \equiv I_{-}+I_{+}$, with
$I_{\mp} \equiv \eta_L\gamma\gamma_c/\Gamma_c(1\mp G)[(1\mp G)\Gamma +
\Gamma_c]$.

\subsection{Throughput and Fidelity Calculations}
To calculate the throughput and fidelity of our singlet-based quantum
communication system, we need
only use the loaded-cavity state, presented in the previous subsection, to
find the erasure,
success, and error probabilities via,
\begin{eqnarray}
P_{\mbox{\scriptsize erasure}} &=&
\left({_{S_x}}\langle
0|\hat{\rho}_{S_x}|0\rangle{_{S_x}}\right)\left({_{S_y}}\langle
0|\hat{\rho}_{S_y}|0\rangle_{S_y}\right) +
\left({_{I_x}}\langle
0|\hat{\rho}_{I_x}|0\rangle{_{Ix}}\right)\left({_{I_y}}\langle
0|\hat{\rho}_{I_y}|0\rangle_{I_y}\right)
\nonumber \\[.12in] &-&
\left({_{S_x}}\langle  0|_{I_y}\langle
0|\hat{\rho}_{S_xI_y}|0\rangle_{I_y}|0\rangle{_{S_x}}\right)
\left({_{S_y}}\langle  0|_{I_x}\langle
0|\hat{\rho}_{S_yI_x}|0\rangle_{I_x}|0\rangle_{S_y}\right),
\\[.12in]
P_{\mbox{\scriptsize success}} &=&
_{\mbox{\scriptsize SI}}\langle\psi|
\hat{\rho}_{\vec{S}\,\vec{I}\,}
|\psi\rangle_{\mbox{\scriptsize SI}}\quad\mbox{and}\quad
P_{\mbox{\scriptsize error}} = 1 -
P_{\mbox{\scriptsize erase}} - P_{\mbox{\scriptsize success}},
\end{eqnarray}
where
$|\psi\rangle_{\mbox{\scriptsize SI}} \equiv
\left(|1\rangle_{S_x}|1\rangle_{I_y}
|0\rangle_{S_y}|0\rangle_{I_x}
-|0\rangle_{S_x}|0\rangle_{I_y}
|1\rangle_{S_y}|1\rangle_{I_x}\right)
/\sqrt{2}$,
is the singlet state.

In Fig.~5 we have plotted the throughput and loss-limited fidelity for our
quantum communication system
under the following assumptions:  OPAs pumped at 1\% of their oscillation
thresholds ($G^2 = 0.01$);
5\,dB of excess loss in each $P$-to-$M$ block path in Fig.~1; 0.2\,dB/km
loss in each fiber;
$\Gamma_c/\Gamma = 0.5$; and $R = 500$\,kHz memory cycling rate.  We see
from this figure that a
throughput of 200\,pairs/sec can be sustained out to an end-to-end path
length ($2L$) of 50\,km, with a
loss-limited fidelity of 97.5\%.
\begin{figure}[htb]
\begin{center}
\includegraphics[width=.8\hsize]{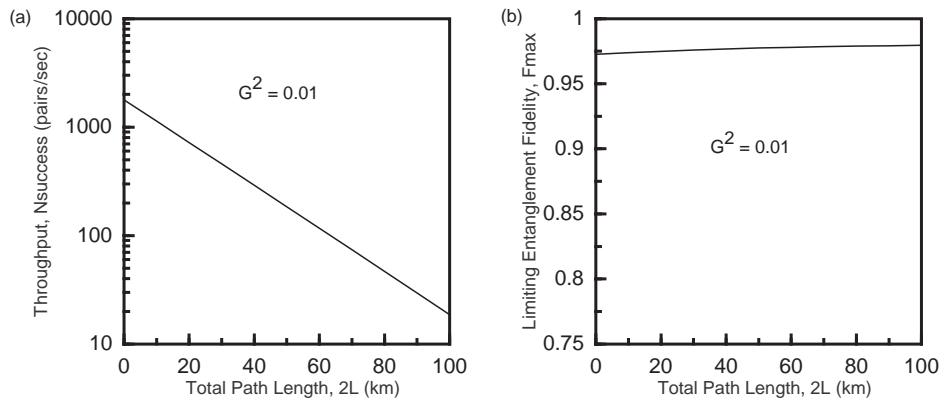}
\end{center}
\vskip -.25cm
\caption{Figures of merit for Fig.~1 configuration.  Left (a), throughput,
$N_{\mbox{\scriptsize success}}$, vs.\ total path length, $2L$. Right (b),
limiting
entanglement fidelity, $F_{\mbox{\scriptsize max}}$, vs.\ total path
length, $2L$. All curves
assume OPAs operating at 1\% of their power thresholds, 5\,dB of excess
loss per $P$-to-$M$
block connection, and 0.2\,dB/km fiber-propagation loss.}
\end{figure}

\section{DISCUSSION}
We have described a single-hop, long-distance, high-fidelity quantum
communication system whose
loss-limited operating range---without entanglement purification or quantum
error
correction---extends well beyond that of previous quantum repeater
proposals.  At $2L = 50$\,km
there is 20\,dB end-to-end loss in our system example, yet, because of the
nondestructive
memory-loading verification, the ultrabright nature of our entanglement
source, and our ability to
employ the low-loss wavelength window in standard telecommunication fiber,
we can sustain appreciable
throughputs and high fidelity.  Of course, this analysis has neglected any
additional degradations
that may arise from residual phase errors that are not common-moded out by
our time-division
multiplexing scheme.  Likewise, the fidelity result we have reported
applies to singlet-based
teleportation assuming perfect Bell-state measurements, and perfect
post-Bell-measurement state
transformation.  Imperfections in any of these areas will reduce the
teleportation fidelity that can
be achieved.  Nevertheless, the Fig.~1 configuration offers substantial
promise for bringing
singlet-based teleportation from a conditional demonstration in the
laboratory to a viable quantum
communication system.

\section{ACKNOWLEDGEMENTS}
This research was supported in part by U.S.\ Army Research Office Grant
DAAD19-00-1-0177.  The author acknowledges fruitful technical discussions
with Phil Hemmer, Prem
Kumar, Seth Lloyd, Selim Shahriar, Franco Wong, and Horace Yuen.